# THE EXTREMELY LOW ACTIVITY COMET 209P/LINEAR DURING ITS EXTRAORDINARY CLOSE APPROACH IN 2014

David G. Schleicher[1] & Matthew M. Knight[1,2]

[1]Lowell Observatory, 1400 W. Mars Hill Rd., Flagstaff, AZ 86001, USA; dgs@lowell.edu
[2]Dept. of Astronomy, University of Maryland, College Park, MD  20742, USA

## ABSTRACT

We present results from our observing campaign of Comet 209P/LINEAR during its exceptionally close approach to Earth during May of 2014, the third smallest perigee of any comet in two centuries. These circumstances permitted us to pursue several studies of this intrinsically faint object, including measurements of gas and dust production rates, searching for coma morphology, and direct detection of the nucleus to measure its properties. Indeed, we successfully measured the lowest water production rates of an intact comet in over 35 years and a corresponding smallest active area, ~0.007 km$^2$. When combined with the nucleus size found from radar (Howell et al. 2014), this also yields the smallest active fraction for any comet, ~0.024%. In all, this strongly suggests that 209P/LINEAR is on its way to becoming an inert object. The nucleus was detected but could not easily be disentangled from the inner coma due to seeing variations and changing spatial scales. Even so, we were able to measure a double-peaked lightcurve consistent with the shorter of two viable rotational periods found by Hergenrother (2014). Radial profiles of the dust coma are quite steep, similar to that observed for some other very anemic comets and suggestive that vaporizing icy grains are present.

1. INTRODUCTION

Although over 550 Jupiter-family comets are now known, only a relatively small fraction of them are also classified as Near-Earth Objects (NEOs), i.e. having perihelia less than 1.3 AU. Of these 59 cases, many have been known for more than a half-century, but another sub-set have only been discovered in the past few decades with the advent of automated search programs such as Spacewatch, LONEOS, NEAT, Catalina Sky Survey, LINEAR, Pan-STARRS, and NEOWISE. Not surprisingly, these newer members have tended to be intrinsically fainter objects, thereby allowing them to previously escape detection. Less clear is whether this faintness is due to small nuclear size, low activity, or a combination of the two. Since many Jupiter-family comets have only small fractions of their surfaces still active (cf. A'Hearn et al. 1995), it is apparent that one end state for a comet is for activity to decrease until it is "dead", potentially resulting in a misidentification as an asteroid rather than a comet. Intrinsically faint objects are inherently difficult to study, thus requiring relatively close encounters with the Earth, and the best recent opportunity took place in 2014 May when Comet 209P/LINEAR (2004 CB = 2008 X2) approached to within 0.056 AU, the closest comet since IRAS-Araki-Alcock (1983d = 1983 H1; 0.031 AU) in 1983 and the third closest in more than two hundred years.

Discovered in 2004 with an asteroidal appearance, 209P/LINEAR was determined to be a comet when a tail was discovered a month later (McNaught 2004). It currently has a 5.1 year orbital period and a perihelion distance of 0.969 AU. Every perihelion passage beginning in 1989 has not only brought LINEAR to within 0.7 AU of Earth but successively ~0.1 AU closer each time, with a perigee of 0.26 AU in 2009. Given multiple favorable apparitions in the past quarter century, and a quite good opportunity back in 1954 (0.41 AU), its "late" discovery was suggestive of quite low activity. The extremely close passage by Earth in 2014 provided an opportunity for radar observations (Howell et al. 2014), while predictions were also made for a new meteor shower or even storm due to the Earth probably crossing a dust trail composed of grains released in the early 1800s (Jenniskens 2006; Ye & Wiegert 2014). Fortunately, in addition to the very small perigee, LINEAR would be well placed for northern hemisphere observers.

Our own planned studies had several components, including gas production rates, jet morphology, and nucleus lightcurve. The comet had simply been too faint to attempt our standard photometric measurement of gas at the two most recent apparitions, so we knew this would be our only opportunity. Gas and/or dust jets have been detected for most highly evolved comets, and the rapid change in viewing geometry that takes place on a close passage could potentially yield a 3-D "image" of LINEAR, similar to our prior work on Hyakutake (1996 B2) (Schleicher & Woodney 2003). Low activity also implied that we might successfully isolate the nuclear signal from the coma, providing the opportunity to measure the rotational lightcurve such as we've done for Comets 10P/Tempel 2 (A'Hearn et al. 1989; Knight et al. 2011) and 2P/Encke (Woodney et al. 2007). This combination of science goals necessitated that most nights be devoted to imaging; a minimal number of nights were scheduled for traditional photoelectric photometry, only one of which was clear (Schleicher 2014). Finally, we note that just prior to the bulk of our observations, two possible rotation periods — 10.93 and 21.86 hr — were reported by Hergenrother (2014), thereby allowing us to better plan our own observations. We present our observations here, along with the results from our analyses for each of our planned goals.

## 2. OBSERVATIONS, REDUCTIONS, AND METHODOLOGIES
*2.1 Overview, Instrumentation, and Observations*

As is typical in such cases, the very small perigee of LINEAR in 2014 was accompanied by a quite short interval when it was both bright enough and available sufficiently long at night for our observations, and the moon could be avoided. Additionally, measurements of the nucleus lightcurve for Comet 6P/d'Arrest were a priority until mid-May, and so imaging of LINEAR was restricted to a few snapshots in early May with more extensive observations waiting until late in the month, near perigee and just prior to LINEAR becoming a southern hemisphere target. While traditional photoelectric photometry was attempted in late April, as expected the comet was still too faint, at about 16$^{th}$ magnitude. Only one night of our May photometry run with the Hall 42 in (1.1 m) telescope was clear, while the other two nights were clouded out. Most of the planned imaging was to make use of our 4.3-m Discovery Channel Telescope (DCT), supplemented with the 42 in telescope. Unfortunately, heavy smoke from the "Slide Fire" in Oak Creek Canyon, 40 km to the NW, adversely affected four of the scheduled nights – the first had smoke in the west but was mostly photometric, two nights proved unusable, and one had a combination of smoke and clouds. Clouds also were present on an additional night. As a result, images were only successfully obtained on a total of 5 nights, several with interruptions, significantly limiting our planned rotational studies. Observing circumstances for both the imaging and the photoelectric photometry are given for usable nights in Table 1, along with the conditions on each night.

[TABLE 1 HERE: OBSERVATIONAL CIRCUMSTANCES]

The CCDs and the photometer characteristics have previously been described (Knight & Schleicher 2015; Schleicher & Bair 2011) and so are not repeated here. Our standard set of HB comet filters was employed with the photometer, isolating the OH, NH, CN, $C_3$, and $C_2$ emission bands along with UV, blue, and green continuum points (Farnham et al. 2000). Due to the faintness of the comet, only a few of these filters were attempted with the imaging, along with a broadband Kron-Cousins R filter (see Table 1).

*2.2 CCD Reductions and Analyses*

While the basic CCD reductions followed standard techniques for bias removal and flat fielding, our multiple goals required a mix of extractions and/or enhancements. The comet was strongly centrally condensed, making it relatively easy to centroid on the nominal nucleus position. We followed our normal methodology (cf. Knight and Schleicher 2015) to enhance images in order to reveal faint underlying structures in the coma, and to conduct aperture photometry for lightcurve analysis and computation of $A(\theta)f\rho$, a proxy for dust production (A'Hearn et al. 1984). We did not attempt to determine gas production rates from images, as the signal-to-noise per pixel was insufficient; the results would have been dominated by uncertainties from bias variations and read-out noise.

Photometric fluxes were measured in a series of apertures centered on the nucleus. The primary apertures used in our analyses reported here had radii fixed in size at the comet of 312 km and 3980 km on all nights. The large aperture size was chosen to match the preferred photometer aperture (see Section 2.3) when measuring $A(\theta)f\rho$ from the images. The smaller aperture was used for the lightcurve analysis during late May. Its size was dictated by the smallest useful extraction aperture on May 21, when the geocentric distance, $\Delta$, was largest. By using an aperture with a fixed physical scale at the comet, all extractions should have contained approximately the same amount of coma contamination since the overall activity of the comet did not change significantly during this interval. We investigated several methods of coma removal in an attempt to isolate the nucleus contribution: extrapolation of the coma in to the nucleus (as used in Knight et al. 2011, 2012), deconvolution of the images with the point spread function then extrapolation of the coma in to the nucleus, and modeling the nucleus/coma contributions by creating synthetic images. We found that the nucleus was approximately 50-70% of the signal in our photometric aperture on all nights, but could not be cleanly isolated due to changes in observing conditions including seeing, the presence of a sunward feature that may have varied in brightness (discussed in Section 3.1), and possible diurnal variations.

As an independent check on our attempts to disentangle the nucleus from the inner-most coma for our small aperture extractions, we used results both from radar observations in late May and brightness measurements in February and March to compute the nominal average brightness for each of our nights' data in May. LINEAR came sufficiently close and had a suitable declination to be studied by radar with the Arecibo telescope and Howell et al. (2014) directly measured an elongated body about 3.9×2.7×2.6 km in size, for an effective radius of ~1.53 km. Using this size, a linear phase function of 0.042 magnitudes per degree (Hergenrother 2014), an assumed geometric albedo of 0.04, and the geometry at the times of our observations (given in Table 1), we estimated the average brightness of the nucleus each night. When compared with the average magnitude in our small photometric aperture each night this implied nucleus fractions of 52-69%, in excellent agreement with our estimates from various coma removal techniques of 50-70% given above. In spite of this confirmation, we ultimately did not attempt to remove the coma contribution from the lightcurve because the coma brightness was not expected to be completely constant with time and the nucleus brightness would certainly be changing with rotation; thus the percentage of coma contamination would be varying both from night to night and during each night, and an incorrect decontamination of the lightcurve would be worse than no adjustment.

Following the same procedure, we next determined the degree to which the nucleus contaminated the large aperture dust measurements. Because most nights exhibited little to no variation in this aperture, mean values for the total extracted flux were computed for each night along with our estimated nucleus contribution. Overall, the nucleus had a small, but non-negligible, contribution – between 17 and 28% – to the total fluxes. As one goal was to look for rotational and seasonal trends, we computed both unadjusted and decontaminated values of $A(\theta)f\rho$, given later in Table 2.

We determined absolute calibrations on photometric nights using broadband standard stars (Landolt 2009, 2013). Typical calibration coefficients were applied to non-photometric nights to allow us to assess the quality of the data. We normalized the lightcurves to the observing geometry on May 30, when the comet was at its closest to Earth, by assuming an inverse square law for $r_H$ and $\Delta$. Due to the varying contributions of nucleus and coma, which have differing phase angle dependencies, we did not correct the lightcurves for phase angle (this will be revisited in Section 3.3). Instead, we applied a nominal nightly offset to bring all nights in approximate agreement by using the overlaps in rotational phase (based on Hergenrother's (2014) rotation period). We excluded portions of the lightcurve on May 27 and 28 where extinction from clouds/smoke exceeded ~0.1 mag; the comet's motion was too fast on these nights to correct the photometry with on chip comparison stars, so the nominal nightly offset also accounts for the non-photometric conditions. For the included data, we estimate the point-to-point uncertainties to be comparable on all nights at about 0.02-0.03 mag.

*2.3 Photometer Reductions and Analyses*

Three complete sets of our traditional narrowband photoelectric photometry were obtained on May 19, only 13 days following perihelion and 10 days prior to perigee. Standard techniques and reduction coefficients (cf. A'Hearn et al. 1995; Schleicher & Bair 2011) were used to compute fluxes, abundances within the photometer entrance aperture, $M(\rho)$, and production rates, $Q$, for each of the five daughter gas species. Water production rates, based on $Q$(OH), were also derived. One-sigma uncertainties were computed based on the observational photon statistics. Dust fluxes for each continuum filter – ultraviolet, blue, and green – were computed along with $A(\theta)f\rho$. As with the imaging extractions, these dust values were contaminated by the nucleus signal but no information regarding the nucleus was available and therefore we used our contamination estimate of 17% from the imaging on May 21 to make a first order correction. (Note that the gas results are unaffected by the nucleus, since continuum removal – whether from dust grains, the nucleus, or a combination of the two – is performed prior to computing gas fluxes.) Because of the comet's close approach to Earth, the phase angle changed significantly and we also normalized all of the dust results to 0° phase angle using a composite dust phase curve (see Schleicher and Bair 2011). Nightly $A(\theta=0°)f\rho$ values are given later in Table 2.

## 3. RESULTS
### *3.1 Coma Morphology*

Comet 209P/LINEAR's overall appearance was much as expected: an obvious dust tail in the anti-solar direction and a faint coma surrounding a strong central condensation associated with the nucleus itself. We applied various image enhancement techniques (e.g., Schleicher and Farnham 2004, Samarasinha and Larson 2014) to look for faint underlying features in the coma. These revealed a small (extending 500-700 km) fan-like feature in the sunward direction in both the R and CN images. The signal-to-noise was insufficient to decontaminate CN images, but since the sunward feature appears in both (along with the dust tail), we suspect its existence in the CN images is simply due to dust contamination. Original and enhanced R and CN images are shown in Figure 1.

[FIGURE 1 HERE: SAMPLE IMAGES]

The sunward feature did not exhibit obvious motion or change in morphology during a night, but showed some variability in position angle (PA) and shape from night to night. One possible interpretation is that it was caused by activity from a near-polar source region and its changing PA from night to night was due simply to changing viewing geometry. If so, the central PA of the feature could be used to constrain the rotational pole (e.g., Knight et al. 2012), however, this did not yield a consistent solution. Modeling to explore other scenarios was beyond the scope of this paper given the limited quantity of high quality imaging data. By combining our data with other datasets, such as the unpublished Arecibo observations by Howell and collaborators, it may yet be possible to constrain the rotation pole and latitude of the source region.

While many, if not most, comets exhibit a mean radial profile for the dust coma that is steeper than the canonical $1/\rho$ profile (cf. Baum et al. 1992), the slope we measured for LINEAR was even steeper than usual. The mean profile from each night is shown in Figure 2. Normalization of each night's data at its peak brightness yields the large relative offsets due to very different effective seeing both between the two telescopes and as a function of the changing geocentric distance of the comet. As is evident, once one is beyond the central seeing region, all of the nights show very similar profiles, having log-log slopes of between -1.9 and -2.1. Since this steep slope begins as soon as seeing effects no longer dominate and before radiation pressure is expected to have a significant effect (i.e. many thousands of kilometers), we conclude that we have strongly "fading" grains similar to the most extreme cases – 28P/Neujmin 1 (-2.0) and 49P/Arend-Rigaux (-1.7) – found by Baum et al. While it remains unclear two decades later whether these are grains shrinking in size, darkening, or some combination, there is almost certainly an icy component to the "dust" grains that is vaporizing away with time.

[FIGURE 2 HERE: MEAN DUST PROFILES; *designed to fit in a single column of the journal*]

### 3.2 Coma Production Rates

Continuing with our continuum data, we next present our *Afρ* results. With the strong departure from a 1/$\rho$ profile, the resulting *Afρ* values would exhibit a very strong trend with aperture size. As previously mentioned we, therefore, used a constant projected aperture size for the extraction for all nights, based on the smaller of the two fixed aperture sizes used with the photometer on May 19, $\rho$ = 3980 km. Results are given in Table 2, where the unadjusted mean values, $A(\theta)f\rho$, for each night are given first. Note that, due to lack of sensitivity in the red, no red measurement is made with the photometer and instead we give the blue continuum value; within the uncertainties, no obvious color trend is evident on the 19th though cometary dust is often reddened and *Afρ* is often somewhat lower at bluer wavelengths.

[TABLE 2 HERE -- AFRHO SUMMARY; *designed to fit in a single column of the journal*]

As discussed in Section 2, the nucleus provides a non-negligible contamination, of about 17-28%, to these coma extractions with the specific value varying both with phase angle due to greatly differing phase effects for the solid nucleus and for dust grains, and with possible coma brightness variations. Our best estimates of the percentage contaminations by the nucleus for each night are also presented in Table 2, along with a decontaminated value and the final phase corrected value to 0° phase angle. The decontaminated, normalized results reveal that the dust coma varied by just over a factor of two during this four-week interval but with no pattern discernable; in fact, the largest normalized value, on May 30, coincides with a minimum in the "nucleus" lightcurve. It remains unclear if the coma variability is associated with rotation or sporadic activity. For comparison, Ye and Wiegert (2014) unexpectedly found that *Afρ* remained near a value of 1 cm on each of four nights within an 8 month interval surrounding perihelion during the comet's 2009 apparition, and did not exhibit an expected drop-off with heliocentric distance. Note that while we do not attempt to convert our *Afρ* results to physical mass loss rates, Ishiguro et al. (2015) and Ye et al. (2016) have performed extensive dust tail analyses to derive these and other properties of the dust grains.

Switching to the gas results, we have given the reduced narrowband fluxes and associated aperture abundances (as logarithms) in Table 3, while resulting log production rates are listed in Table 4. Since the uncertainties are unbalanced in log-space, though the upper and lower values are equal in percentage, we list only the upper or "+" log sigmas in Table 4; the lower or "-" values can be readily calculated. The vectorial-equivalent water production rate is also given in the right-most column. It is readily evident that the measurements for each species are self-consistent within the photometric uncertainties among the three observational sets. We have therefore computed the mean values for the night, along with each sigma of the mean and included these values at the bottom of Table 4. Looking first at the resulting gas production rate ratios, we find that LINEAR falls within our "typical" compositional class, though near the low end for the $C_3$-to-CN production rate ratio within the typical group (Schleicher & Bair 2014). A typical classification for a Jupiter-family comet is not unusual, since while the majority of comets showing depletion in one or more species are Jupiter-family objects, more than half of Jupiter-family comets have typical composition. Finally, the dust-to-gas ratio, based on $Af\rho/Q$(OH), is about a factor of two higher than average but not at all unusual.

[TABLE 3 HERE: NARROWBAND FLUXES]

[TABLE 4 HERE: NARROWBAND PRODUCTION RATES]

LINEAR's absolute water production rate, however, was a surprise. While we expected it to be relatively low, for the reasons mentioned in the Introduction, the measured value of 2.5±0.2 × $10^{25}$ molecules s$^{-1}$ was below what we have previously measured for *any* intact or whole comet since our composition database program began over 35 years ago. Our next lowest water production was for 10P/Tempel 2 more than 100 days before perihelion and prior to its seasonal "turn-on" of activity (see Knight et al. 2012), but its mean apparitional value is much larger. Several other comets, including 28P/Neujmin 1 and 45P/Honda-Mrkos-Pajdusakova (H-M-P), have successfully been measured by us to have $Q$(H$_2$O) as low as ~ 2×10$^{26}$ molecules s$^{-1}$ (Schleicher & Bair 2014), still about 8× larger than LINEAR. Even relatively small component "G" from 73P/Schwassmann-Wachmann 3's breakup was comparable to Neujmin 1 and, in fact, only tiny fragment "R" from S-W 3 (Schleicher & Bair 2011) had a slightly smaller successful water measurement than that of LINEAR.

An alternative to comparing the absolute water production rates is to compare the effective active areas required to produce the amount of water in the coma from vaporization of ices at the surface of the nucleus. We continue to use a water vaporization model based on Cowan & A'Hearn (1979), particularly an updated rapid rotator pole-on to the Sun that yields the maximum rate of vaporization per unit area. While different assumptions for the vaporization model, such as the isothermal case, would yield somewhat larger area estimates, we have chosen to use the identical scenario here as we have for all of the other comets in our database. As with the water production, LINEAR is our new extreme case for an intact comet, with an active area of only 0.007 km$^2$. Note that if our conclusion from Section 3.1 is correct and that icy grains are needed to explain the steep radial profiles of the "dust" grains, then the sublimation of the water ice component would supply a portion of the observed water vapor and hence an even smaller active area would be needed than what we've computed. For comparison, and averaged for a given comet over all of our observations, H-M-P was our previous record holder with a median value of 0.35 km$^2$, about 50× higher than LINEAR! Having only measured a water value on one night, a reasonable question is whether LINEAR experiences a strong seasonal effect and our observations were atypically small? An examination of visible magnitudes compiled by S. Yoshida[1] in 2004, 2009, and 2014 suggests just the opposite, with brightnesses after perihelion higher than would be expected if outgassing were symmetric about perihelion. Dust coma measurements in 2014 by Ishiguro et al. (2015) also indicate maximum activity occurred at or slightly after perihelion. Both results imply that an average active area obtained over several months would likely have been even smaller than our measurement since our data were acquired just 13 days after perihelion.

If the size of the nucleus is known, one can also compute the fractional active area. While in many cases the sizes are based on nucleus extracted magnitudes and an assumed albedo, as already discussed in Section 2.2 LINEAR came sufficiently close and had a suitable declination to be studied by radar with the Arecibo telescope. Using Howell et al.'s (2014) measurements, we compute an effective radius of ~1.53 km and a surface area of ~29 km$^2$. When combined with the previously determined effective active area, we obtain a fractional active area of ~0.024%, again below that of any other comet we've successfully observed in more than three decades. In comparison, our prior smallest fraction was 0.05% for Neujmin 1, while H-M-P was 4%. To better place LINEAR with other comets in our database, we plot in log-log space in Figure 3 the active fractions and the active areas. Note that there is a strong observational bias against determining the nucleus size of long-period comets due to the usual presence of a significant coma, thereby making it very difficult to separate out the nucleus signal; in fact, only five long-period comets in our database have nuclear size estimates and associated active fractions. This very problem of nucleus detectability for long-period comets, however, directly implies a relatively large active fraction must exist. In conclusion, 209P/LINEAR is truly an extreme case regarding both its active area and active fraction.

[FIGURE 3 HERE: ACTIVE AREAS AND FRACTIONAL AREAS; *designed to fit in a single column*]

*3.3 Nucleus Lightcurve Results*

Another goal of this project was to measure a rotational lightcurve of the nucleus and determine its period. We succeeded in detecting the nucleus and obtaining a partial lightcurve, but as noted in Section 2.2, we were unable to reliably disentangle the nucleus signal from the inner coma contribution of the lightcurve and instead used a small fixed projected aperture size, 312 km, for extractions, thereby hoping to minimize effects to the lightcurve from the coma. These values are given in Table 5. Additionally, too many scheduled nights of observations were lost to clouds and smoke to obtain sufficient temporal coverage to perform a period determination. We therefore instead utilized the possible rotational periods announced by Hergenrother (2014), 10.93 and 21.86 hr, and phased our own data to these values. These solutions are shown in Figure 4. As can be seen, our measurements are completely consistent with and give a very sensible phased lightcurve with either value.

---

[1] http://www.aerith.net/index.html

[TABLE 5 – LIGHTCURVE DATA; *designed to fit on a single page, as given*]

[FIGURE 4 – PHASED LIGHTCURVE; 2 PANELS (BOTH PERIODS)]

We also tried slightly varying the periods to see if we could fine-tune the answer. Given the uncertainties on the absolute calibration from night to night (see Section 2.2), there is some suggestion that the best solution might be 0.01-0.02 less than Hergenrother's (2014) values but we do not have confidence in this finding given possible changes in the coma contribution. As to which period is correct, Hergenrother preferred the long value because phasing his data (from February and March 2014) showed a double-peaked curve for his long value but only one strong peak in 10.93 hr phase plot. In comparison, for our data it is easy to infer a double-peaked curve from the short period solution, and three or four peaks for the longer period. Radar provides the tie-breaker, and Howell et al. (2014) find rotational velocities are too fast for the longer period; thus we are confident that LINEAR's rotation period must be the shorter of the two options, i.e. very close to 10.93 hr.

The nucleus dimensions estimated by radar imply a peak-to-trough lightcurve amplitude of ~0.4 mag, which is approximately equal to the maximum amplitude of the lightcurve that we saw on any individual night, but is less than the amplitude of 0.6-0.7 that we infer by arbitrarily offsetting each night's lightcurve in order to produce a lightcurve that appears smooth by eye. There are several possible explanations for this discrepancy. First, Howell (private communication) notes that the nucleus could be even more elongated than they reported, in which case the amplitude would be higher. Second, the comet's activity could vary diurnally, increasing the amplitude. Third, our nightly offsets could be in error, causing us to over estimate the amplitude. Another consideration is that our lightcurve shape differs significantly from Hergenrother's (personal communication), where his 10.93 hr solution had only one strong peak. As his February data showed no evidence of coma, this implies either a very irregular shaped nucleus combined with a significant change in sub-solar latitudes between Feb/Mar and May or, given the radar results, perhaps a nucleus lightcurve having albedo variations across the surface; note that variations of only a percent from the nominal 4% value would be significant and well within the range of values observed on several nuclei by spacecraft. We favor a combination of shape/viewing geometry and albedo effects, but it is unclear, even with the synergistic characteristics of these data sets, whether a definitive solution can ultimately be obtained.

4. DISCUSSION AND SUMMARY

As hoped, Comet 209P/LINEAR indeed proved to be both an interesting and extreme object. In spite of a much more limited data set than planned due to adverse weather conditions coupled with smoke from a major forest fire, we were able to discern a number of important results. First we confirmed our expectation that it was only discovered in this century because of its low intrinsic activity level, requiring an active area on the surface of the nucleus – about 0.007 km$^2$ or the size of a football field – less than for any other comet we've measured in nearly four decades. Moreover, 209P/LINEAR proved to have an unusually large nucleus size given its low level of activity, implying the smallest active fraction of any "live" comet of which we are aware. The extremely low water production measured, the very small active area associated with the release of that amount of water vapor, and the miniscule active fraction for Comet 209P/LINEAR, are all strong indicators of a highly evolved object very near the end of its "cometary" life. We can easily imagine that in a few more centuries or millennia it would simply appear asteroidal with no sign of its cometary origin other than a possible dust trail. Based on their own 2014 dust coma measurements along with meteor shower results, Ye et al. (2016) have recently arrived at the same conclusion. In fact, several asteroids having "comet-like orbits" have also been identified with meteor streams or dust trails, including 2201 Oljato and 3200 Phaethon, while 4015 Wilson-Harrington once had a tail and was given a dual designation. While attempts to obtain unambiguous detections of either dust comae or gas emission have been unsuccessful, the gas searches have generally had less sensitivity than our 209P/LINEAR efforts. For instance, Chamberlin et al. (1996) place 3-sigma upper limits for CN emission of these three NEOs corresponding to production rates of $1\times10^{23}$ mol s$^{-1}$ for Oljato and Phaethon, and $4\times10^{22}$ mol s$^{-1}$ for Wilson-Harrington, while we measured a value for CN of $6\times10^{22}$ mol s$^{-1}$ for LINEAR under more favorable circumstances. Detections of an occasional tail and perihelion brightening of Phaethon are also very suggestive of episodic dust release (cf. Jewitt et al. 2013). Finally, similar to some other very anemic comets, the dust coma's radial profile of LINEAR is very steep, likely due to grains shrinking or darkening with time and thus having an icy component (Baum et al. 1992). It is unclear, however, why such grains should dominate the coma of highly evolved comets.

A related issue regarding the death process in comets is whether very low levels of activity are associated with one or at most a few very small, isolated source regions or are instead "leakage" over a much larger fraction of a nucleus' surface. In the case of 209P/LINEAR, the short, faint dust feature is suggestive of arising from an isolated source, as is limited evidence for seasonal variations based on brightnesses from 2004, 2009, and 2014 (Yoshida[1]) that imply that the comet is somewhat brighter after perihelion than before. Also, Hergenrother (2014) and Ishiguro et al. (2015) claim that activity only turns on in-bound near 1.4 AU, implying that a source region is only then coming into sunlight. We conclude that this combination of behavior suggests an isolated source, but more near-death comets should be investigated. With even more sensitive surveys arriving we anticipate more close-approach opportunities in the future. In fact, the next opportunity takes place in 2016 March/April when Comet 252P/LINEAR comes to within 0.04 AU of Earth.

Comet LINEAR's typical gas composition provides yet another example that evolution is *not* the cause of why many Jupiter-family comets exhibit strong carbon-chain depletion. We also see no evidence that a canonical value of 4% for the albedo of the nucleus is incorrect; indeed, combining this average albedo and standard solid body phase angle effects for the nucleus, along with standard dust phase effects for the coma yield self-consistent results from our own data, the observations from Hergenrother (2014) a few months earlier, and the radar measurements by Howell et al. (2014). Though we had an insufficient number of clear nights to obtain full rotational coverage and uniquely determine the comet's precise rotation period, our data are also fully consistent with the two possible solutions found by Hergenrother. Our observations did yield clear evidence of a change in the lightcurve shape from February to May, and suggests that either the shape of the nucleus deviates significantly from a simple ellipsoid and/or there are albedo variations across the surface; subsequent analyses may able to distinguish among these and other possibilities such as rotationally induced brightness changes of the coma.

Unfortunately, the next several perihelion passages of 209P/LINEAR will progressively move away from Earth, yielding much less favorable circumstances. Even so, perihelion will remain near 0.97 AU and the comet's activity levels will likely remain extremely low, thereby making LINEAR a potentially viable safe and energetically relatively easy-to-get-to target for an NEO manned missions in the future. Given this possibility and the likelihood that we are seeing a comet very close to its "death", additional observations at subsequent apparitions are warranted.

## ACKNOWLEDGEMENTS

We thank E. Howell for discussions regarding the radar results and C. Hergenrother for making available his lightcurves prior to publication. We thank A. Bair for comparisons with the compositional database and assistance with the creation of several tables. We also thank our DCT telescope operators, T. Pugh and H. Larson. These results made use of Lowell Observatory's Discovery Channel Telescope, supported by Lowell, Discovery Communications, Boston University, the University of Maryland, the University of Toledo, Northern Arizona University, and Yale University. This research was supported by NASA's Planetary Astronomy Program (grant NNX14AG81G).

# REFERENCES

A'Hearn, M. F., Millis, R. L., Schleicher, D. G., Osip, D. J., & Birch, P. V. 1995. Icarus 118, 223.

A'Hearn, M. F., Campins, H., Schleicher, D. G., & Millis, R. L. 1989. ApJ 347, 1155.

A'Hearn, M. F., Schleicher, D. G., Feldman, P. D., Millis, R. L., & Thompson, D. T. 1984. AJ 89, 579.

Baum, W. A., Kreidl, T. J., & Schleicher, D. G. 1992. AJ 104, 1216.

Chamberlin, A. B., McFadden, L.-A., Schulz, R., Schleicher, D. G., & Bus, S. J. 1996. Icarus 119, 173.

Cowan, J. J., & A'Hearn, M. F. 1979. Moon Planets 21, 155.

Farnham, T. L., Schleicher, D. G., & A'Hearn, M. F. 2000. Icarus 147, 180.

Hergenrother, C. 2014. IAU CBET 3870.

Howell, E. S., Nolan, M. C., Taylor, P. A., et al. 2014. In AAS/DPS Meeting Abstracts 46, #209.24.

Ishiguro, M., Kuroda, D., Hanayama, H., et al. 2015. ApJ Letters 798, #L34.

Jenniskens P. (2006). Meteor Showers and their Parent Comets. Cambridge University Press, 790 pages.

Jewitt, D., Li, J., & Agarwal, J. 2013. ApJ Letters 771, #L36.

Knight, M. M., Farnham, T. L., Schleicher, D. G., & Schwieterman, E. W. 2011. AJ 141, #2.

Knight, M. M., Schleicher, D. G., Farnham, T. L., Schwieterman, E. W., & Christensen, S. R. 2012. AJ 144, #153.

Knight, M. M., & Schleicher, D. G. 2015. AJ 149, #19.

Landolt, A.U. 2009. AJ 137, 4186.

Landolt, A.U. 2013. AJ 146, 131.

McNaught, R. H. 2004. IAU Circular 8314.

Samarasinha, N. H., & Larson, S. M. 2014. Icarus 239, 168.

Schleicher, D. 2014. IAU CBET 3881.

Schleicher, D. G., & Bair, A. N. 2011. AJ 141, 177.

Schleicher, D. G., & Bair, A. N. 2014. Asteroids, Comets, Meteors 2014, Session 4-2-1, Paper #7.

Schleicher, D. G., & Farnham, T. L. 2004. In Comets II (M. C. Festou, H. U. Keller, & H. A. Weaver, eds.) Univ. of Arizona, Tucson. 449-469.

Schleicher, D. G., & Woodney, L. M. 2003. Icarus 162, 190.

Woodney, L. M., Schleicher, D. G., Reetz, K. M., & Ryan, K. J. 2007. BAAS 39, 486.

Ye, Q., & Wiegert, P. A. 2014. MNRAS 437, 3283.

Ye, Q.-Z.,Hui, M.-T., Brown, P. G., et al. 2016. Icarus 264, 48.

FIGURE CAPTIONS

**Figure 1.** Representative images of 209P/LINEAR from each successful night in 2014 May. All images in a column are from the same night, and the date is given in the top row. The bottom row shows unenhanced CN images, the middle row shows unenhanced R images, and the top row shows the same R images enhanced by azimuthal median subtraction. All panels have the same physical scale as projected at the comet; a scale bar is shown in the right-most column. Each image is centered on the nucleus; the position of the center is indicated by the white bars on the bottom row. The unenhanced images are strongly centrally condensed, but have been stretched so the coma and tail can be seen; in these frames white is bright and blue or black is faint. The enhanced images have been stretched to best show the sunward feature; red is bright and blue is faint. Trailed stars are visible in some CN images but not in the R frames because the R frames are median stacks of many more images during the night, allowing better removal of the background stars. The CN frames have not had the continuum removed, as several nights were not photometric, and the tail remains obvious.

**Figure 2.** Representative dust profiles are shown. Once beyond the seeing disk, which differs with telescope and the changing geocentric distance, the profile is significantly steeper than $1/\rho$, with a log-log slope of -1.9 to -2.1. The vertical lines at the top indicate the range of distances over which the slope was measured for each night's profile. Similarly steep profiles were also seen for some other very anemic comets (cf. Baum et al. 1992).

**Figure 3.** Log of the fractional active area of the nucleus plotted as a function of the log of the effective active area needed to produce the observed production rate of water. All comets having both a water measurement and a value for the nucleus size are included from our photometric database (Schleicher & Bair 2014). Symbols are based on the dynamical classification, with Jupiter-family comets as triangles, Halley-type comets as squares, and the few long-period comets that have a size estimate shown as circles. 209P/LINEAR has a somewhat smaller active fraction (0.024%) than any other comet (Neujmin 1 is next smallest, followed Halley-type objects LONEOS (2001 OG10) and P/Siding Spring (2006 HR30)), and LINEAR has a much smaller active area (0.007 km$^2$) than the next smallest, Honda-Mrkos-Pajdusakova. A few comets at the top of the plot have active fractions exceeding 100%, and are thought to have some or most of their outgassing from icy grains within their comae.

**Figure 4.** 2014 May lightcurve data phased to a period of 10.93 hr (left panel) and 21.86 hr (right panel). These period solutions are based on lightcurve data obtained between Feb 10 and Mar 10 by Hergenrother (2014); no adjustments for changes in synodic effects are made here but such changes should be small. As evident, our data are consistent with both solutions but cannot distinguish between them. Note that Howell et al. (2014) rule out the longer solution based on incompatibility with Arecibo radar observations.

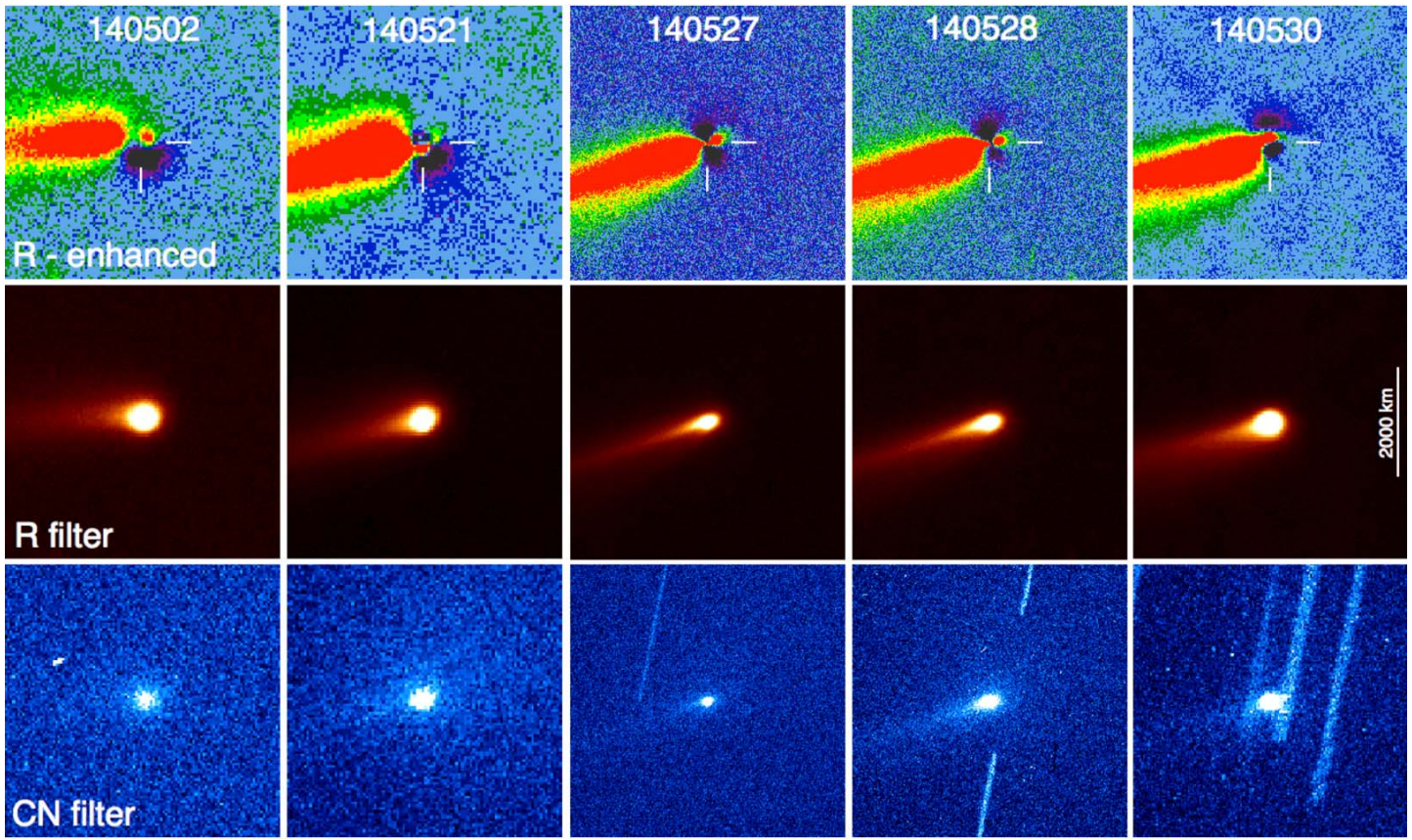

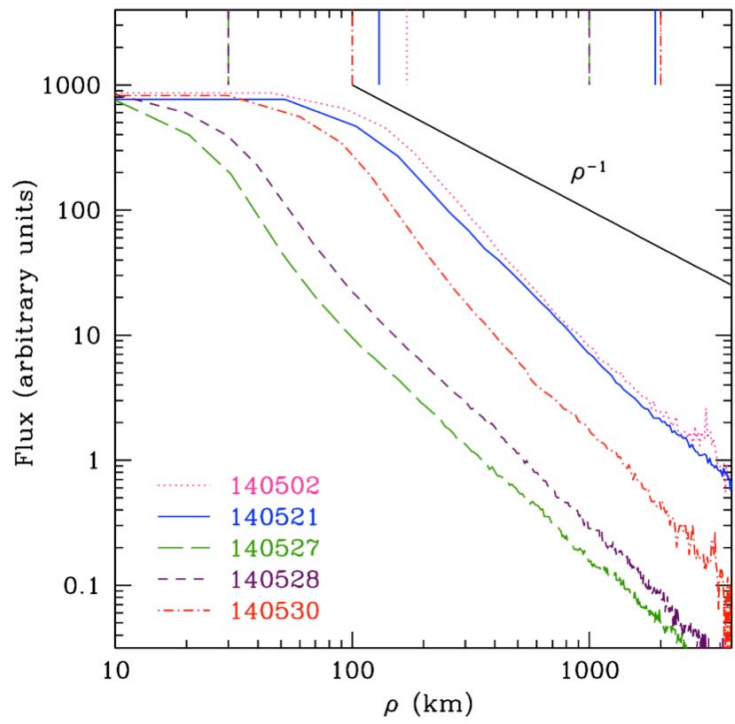

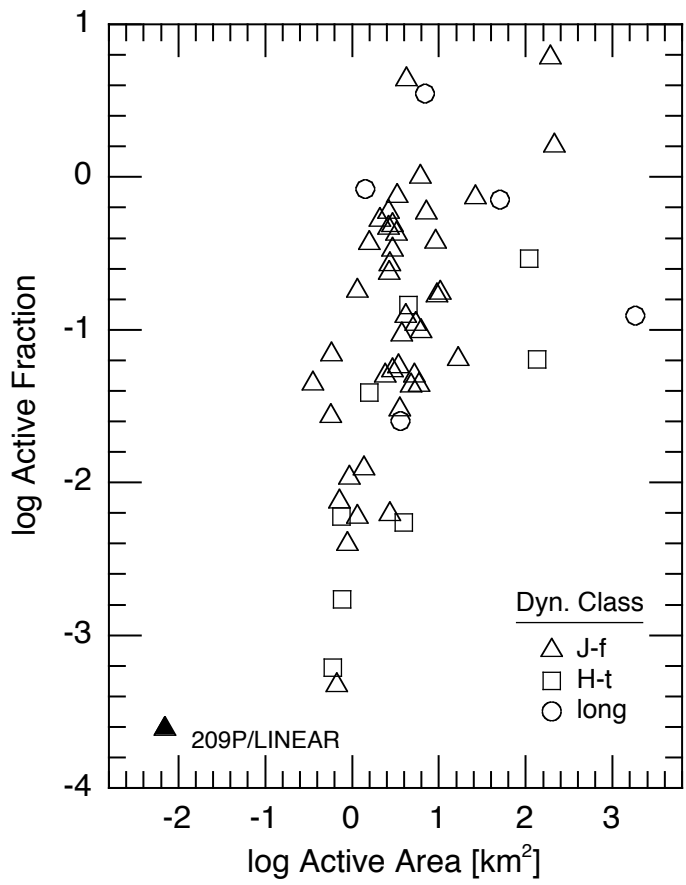

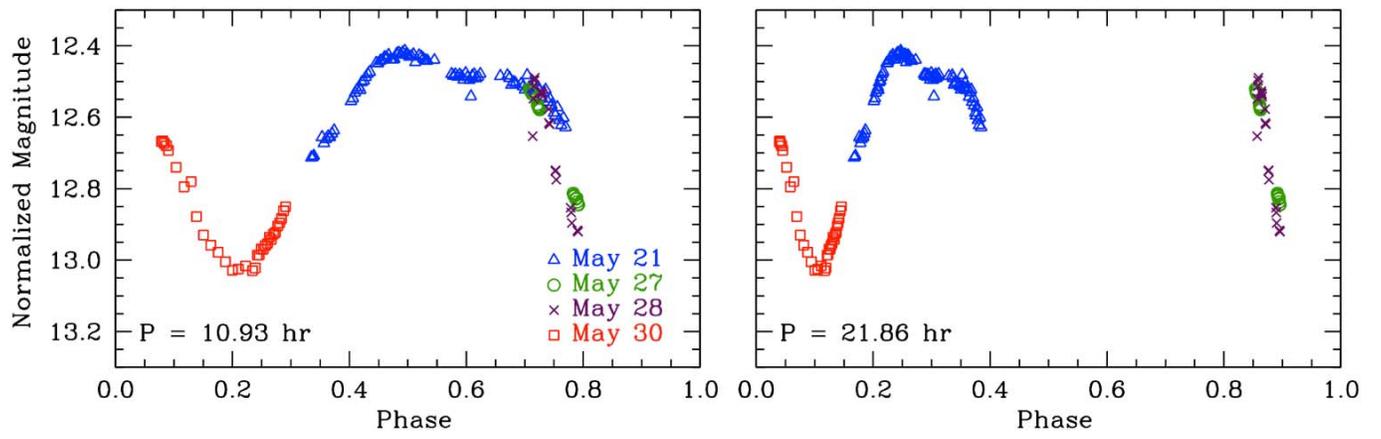

**Table 1**
Observing Circumstances for Comet 209P/LINEAR[a]

| UT Date | UT Range | Tel | Filters | $r_H$ (AU) | $\Delta$ | Phase (°) | Phase Adj[b] | PA of Sun (°) | Conditions |
|---|---|---|---|---|---|---|---|---|---|
| May 2 | 4:44-4:52 | DCT | R, CN | 0.971 | 0.263 | 90.4 | +0.27 | 272.5 | Photometric |
| May 19 | 4:25-5:04 | 42in | All narrowband | 0.987 | 0.113 | 99.5 | +0.16 | — | Photometric |
| May 21 | 3:39-5:38 | 42in | R, CN, BC | 0.992 | 0.097 | 99.0 | +0.17 | 284.2 | Photometric most of time; heavy smoke in west |
| May 27 | 3:43-6:45 | DCT | R, CN, BC | 1.014 | 0.059 | 87.4 | +0.29 | 291.6 | Clouds, smoke |
| May 28 | 3:45-5:57 | DCT | R, CN, BC | 1.018 | 0.057 | 83.4 | +0.33 | 291.7 | Clouds |
| May 30 | 3:28-5:48 | 42in | R, CN, BC, $C_3$ | 1.027 | 0.056 | 74.6 | +0.40 | 289.4 | Possible cirrus |

[a] All parameters are given for the midpoint of each night's observations; perihelion was at 2014 May 6.3.
[b] Adjustment to 0° phase angle to log($Af\rho$) values based on assumed phase function (see text).

**Table 2**
Dust $Af\rho$ Results for Comet 209P/LINEAR

| UT Date | log($A(\theta)f\rho$)[a] [total] | Nucleus fraction[b] | log($A(\theta)f\rho$)[c] [coma] | log($A(\theta=0°)f\rho$)[d] [coma] |
|---|---|---|---|---|
| May 2 | +0.12±0.02 | 0.19±0.02 | +0.03±0.03 | +0.30±0.03 |
| May 19 | +0.02±0.03 | 0.17±0.02 | –0.06±0.04 | +0.10±0.04 |
| May 21 | +0.03±0.04 | 0.17±0.02 | –0.05±0.05 | +0.12±0.05 |
| May 27 | +0.05±0.02 | 0.25±0.03 | –0.07±0.04 | +0.22±0.04 |
| May 28 | +0.10±0.06 | 0.28±0.03 | –0.04±0.08 | +0.29±0.08 |
| May 30 | +0.22±0.02 | 0.28±0.03 | +0.08±0.04 | +0.48±0.04 |

[a] $\rho$=3980 km for whole table, units of $Af\rho$ are cm. Uncertainty is the standard deviation of all magnitude measurements for the night propagated to log($Af\rho$), which was always larger than the estimated error due to the uncertainty in calibrations.
[b] Fraction of the signal in the $Af\rho$ aperture that is due to the nucleus. Calculated from predicted nucleus brightness based on radar size (see text for assumptions). Uncertainty is assumed to be 10%.
[c] Resulting coma $Af\rho$ after removing the nucleus contribution.
[d] Phase adjusted coma $Af\rho$.

**Table 3**
Narrowband Photometric Fluxes and Aperture Abundances for Comet 209P/LINEAR

| UT Date | Aperture | | log Emission Band Flux (erg cm$^{-2}$ s$^{-1}$) | | | | | log Continuum Flux (erg cm$^{-2}$ s$^{-1}$ Å$^{-1}$) | | | log $M(\rho)$[a] (molecule) | | | | |
|---|---|---|---|---|---|---|---|---|---|---|---|---|---|---|---|
| | Size (arcsec) | log $\rho$ (km) | OH | NH | CN | $C_3$ | $C_2$ | UV | Blue | Green | OH | NH | CN | $C_3$ | $C_2$ |
| May 19.18 | 204.5 | 3.92 | −11.37 | −11.95 | −11.42 | −11.91 | −11.52 | −14.35 | −13.96 | −14.22 | 28.78 | 26.67 | 26.46 | 25.63 | 26.37 |
| May 19.20 | 97.2 | 3.60 | −11.92 | −12.54 | −11.95 | −12.12 | −12.12 | −14.71 | −14.10 | −14.34 | 28.23 | 26.08 | 25.94 | 25.42 | 25.77 |
| May 19.21 | 97.2 | 3.60 | −11.79 | −12.80 | −11.86 | −12.14 | −12.07 | −14.64 | −14.19 | −14.25 | 28.36 | 25.83 | 26.02 | 25.40 | 25.82 |

[a] To compute $M(\rho)$ at the comet's heliocentric velocity of +4.6 km s$^{-1}$, the log of the fluorescence efficiencies ($L/N$), dependent on the velocity, that were used were as follows: -14.595 (OH); -13.070 (NH); and -12.331 (CN).

**Table 4**
Narrowband Photometric Production Rates for Comet 209P/LINEAR

| UT Date | log $\rho$ (km) | log $Q$[a] (molecule s$^{-1}$) | | | | | log $A(\theta)f\rho$[a] (cm) | | | log $Q$ H$_2$O |
|---|---|---|---|---|---|---|---|---|---|---|
| | | OH | NH | CN | $C_3$ | $C_2$ | UV | Blue | Green | |
| May 19.18 | 3.92 | 25.23 .07 | 23.36 .11 | 22.74 .03 | 21.75 .28 | 22.85 .09 | −0.18 .27 | −0.13 .12 | −0.36 .20 | 25.37 |
| May 19.20 | 3.60 | 25.22 .12 | 23.32 .18 | 22.73 .06 | 21.91 .27 | 22.76 .15 | −0.22 .29 | +0.06 .08 | −0.16 .14 | 25.35 |
| May 19.21 | 3.60 | 25.34 .11 | 23.06 .28 | 22.81 .04 | 21.89 .26 | 22.81 .13 | −0.15 .28 | −0.03 .10 | −0.07 .12 | 25.48 |
| Mean | | 25.27 .04 | 23.27 .08 | 22.76 .03 | 21.86 .05 | 22.81 .03 | −0.18 .02 | −0.03 .05 | −0.18 .07 | 25.40 |

[a] Production rates, followed by the upper, i.e., the positive, uncertainty. The "+" and "−" uncertainties are equal as percentages, but unequal in log-space; the "−" values can be computed.

**Table 5**
Inner Coma[a] Lightcurve for Comet 209P/LINEAR[a]

| Date | UT (hr) | m_a | m_c | Date | UT (hr) | m_a | m_c | Date | UT (hr) | m_a | m_c | Date | UT (hr) | m_a | m_c |
|---|---|---|---|---|---|---|---|---|---|---|---|---|---|---|---|
| May 21 | 3.675 | 14.993 | 12.713 | May 21 | 5.689 | 14.707 | 12.427 | May 21 | 8.131 | 14.828 | 12.548 | May 28 | 4.188 | 13.468 | 12.748 |
| May 21 | 3.702 | 14.992 | 12.712 | May 21 | 5.733 | 14.711 | 12.431 | May 21 | 8.168 | 14.847 | 12.567 | May 28 | 4.200 | 13.495 | 12.775 |
| May 21 | 3.722 | 14.988 | 12.708 | May 21 | 5.771 | 14.716 | 12.436 | May 21 | 8.212 | 14.867 | 12.587 | May 28 | 4.466 | 13.573 | 12.853 |
| May 21 | 3.873 | 14.936 | 12.656 | May 21 | 5.808 | 14.721 | 12.441 | May 21 | 8.237 | 14.875 | 12.595 | May 28 | 4.477 | 13.587 | 12.867 |
| May 21 | 3.911 | 14.953 | 12.673 | May 21 | 5.846 | 14.722 | 12.442 | May 21 | 8.275 | 14.852 | 12.572 | May 28 | 4.487 | 13.616 | 12.896 |
| May 21 | 3.951 | 14.941 | 12.661 | May 21 | 5.975 | 14.720 | 12.440 | May 21 | 8.312 | 14.888 | 12.608 | May 28 | 4.589 | 13.637 | 12.917 |
| May 21 | 3.987 | 14.934 | 12.654 | May 21 | 6.296 | 14.759 | 12.479 | May 21 | 8.350 | 14.902 | 12.622 | May 28 | 4.600 | 13.640 | 12.920 |
| May 21 | 4.025 | 14.936 | 12.656 | May 21 | 6.334 | 14.764 | 12.484 | May 21 | 8.388 | 14.882 | 12.602 | May 28 | 4.611 | 13.640 | 12.920 |
| May 21 | 4.062 | 14.926 | 12.646 | May 21 | 6.371 | 14.760 | 12.480 | May 21 | 8.429 | 14.908 | 12.628 | May 30 | 3.470 | 12.668 | 12.668 |
| May 21 | 4.100 | 14.917 | 12.637 | May 21 | 6.409 | 14.767 | 12.487 | May 27 | 5.837 | 13.311 | 12.521 | May 30 | 3.488 | 12.666 | 12.666 |
| May 21 | 4.417 | 14.835 | 12.555 | May 21 | 6.445 | 14.761 | 12.481 | May 27 | 5.849 | 13.311 | 12.521 | May 30 | 3.521 | 12.673 | 12.673 |
| May 21 | 4.455 | 14.827 | 12.547 | May 21 | 6.486 | 14.766 | 12.486 | May 27 | 5.859 | 13.317 | 12.527 | May 30 | 3.560 | 12.680 | 12.680 |
| May 21 | 4.491 | 14.811 | 12.531 | May 21 | 6.523 | 14.775 | 12.495 | May 27 | 5.870 | 13.318 | 12.528 | May 30 | 3.598 | 12.693 | 12.693 |
| May 21 | 4.529 | 14.803 | 12.523 | May 21 | 6.561 | 14.757 | 12.477 | May 27 | 5.881 | 13.324 | 12.534 | May 30 | 3.741 | 12.740 | 12.740 |
| May 21 | 4.567 | 14.796 | 12.516 | May 21 | 6.598 | 14.768 | 12.488 | May 27 | 5.985 | 13.354 | 12.564 | May 30 | 3.888 | 12.795 | 12.795 |
| May 21 | 4.592 | 14.804 | 12.524 | May 21 | 6.636 | 14.776 | 12.496 | May 27 | 5.996 | 13.359 | 12.569 | May 30 | 4.020 | 12.780 | 12.780 |
| May 21 | 4.619 | 14.782 | 12.502 | May 21 | 6.659 | 14.822 | 12.542 | May 27 | 6.007 | 13.363 | 12.573 | May 30 | 4.119 | 12.878 | 12.878 |
| May 21 | 4.655 | 14.778 | 12.498 | May 21 | 6.683 | 14.764 | 12.484 | May 27 | 6.018 | 13.367 | 12.577 | May 30 | 4.254 | 12.930 | 12.930 |
| May 21 | 4.693 | 14.768 | 12.488 | May 21 | 6.721 | 14.760 | 12.480 | May 27 | 6.028 | 13.369 | 12.579 | May 30 | 4.390 | 12.958 | 12.958 |
| May 21 | 4.731 | 14.759 | 12.479 | May 21 | 6.759 | 14.771 | 12.491 | May 27 | 6.658 | 13.603 | 12.813 | May 30 | 4.525 | 12.978 | 12.978 |
| May 21 | 4.768 | 14.753 | 12.473 | May 21 | 6.795 | 14.767 | 12.487 | May 27 | 6.668 | 13.608 | 12.818 | May 30 | 4.666 | 13.005 | 13.005 |
| May 21 | 4.887 | 14.728 | 12.448 | May 21 | 6.833 | 14.759 | 12.479 | May 27 | 6.678 | 13.611 | 12.821 | May 30 | 4.797 | 13.029 | 13.029 |
| May 21 | 4.926 | 14.729 | 12.449 | May 21 | 7.210 | 14.765 | 12.485 | May 27 | 6.700 | 13.618 | 12.828 | May 30 | 4.904 | 13.025 | 13.025 |
| May 21 | 4.964 | 14.722 | 12.442 | May 21 | 7.342 | 14.763 | 12.483 | May 27 | 6.710 | 13.619 | 12.829 | May 30 | 5.038 | 13.016 | 13.016 |
| May 21 | 5.001 | 14.720 | 12.440 | May 21 | 7.380 | 14.770 | 12.490 | May 27 | 6.722 | 13.616 | 12.826 | May 30 | 5.169 | 13.030 | 13.030 |
| May 21 | 5.039 | 14.712 | 12.432 | May 21 | 7.417 | 14.789 | 12.509 | May 27 | 6.732 | 13.627 | 12.837 | May 30 | 5.221 | 13.022 | 13.022 |
| May 21 | 5.080 | 14.712 | 12.432 | May 21 | 7.455 | 14.782 | 12.502 | May 27 | 6.743 | 13.634 | 12.844 | May 30 | 5.257 | 12.986 | 12.986 |
| May 21 | 5.117 | 14.706 | 12.426 | May 21 | 7.491 | 14.784 | 12.504 | May 27 | 6.754 | 13.635 | 12.845 | May 30 | 5.295 | 12.986 | 12.986 |
| May 21 | 5.155 | 14.719 | 12.439 | May 21 | 7.600 | 14.787 | 12.507 | May 28 | 3.757 | 13.373 | 12.653 | May 30 | 5.332 | 12.969 | 12.969 |
| May 21 | 5.191 | 14.719 | 12.439 | May 21 | 7.636 | 14.793 | 12.513 | May 28 | 3.769 | 13.268 | 12.548 | May 30 | 5.370 | 12.970 | 12.970 |
| May 21 | 5.229 | 14.718 | 12.438 | May 21 | 7.674 | 14.802 | 12.522 | May 28 | 3.779 | 13.233 | 12.513 | May 30 | 5.410 | 12.959 | 12.959 |
| May 21 | 5.272 | 14.702 | 12.422 | May 21 | 7.711 | 14.762 | 12.482 | May 28 | 3.790 | 13.217 | 12.497 | May 30 | 5.448 | 12.954 | 12.954 |
| May 21 | 5.309 | 14.700 | 12.420 | May 21 | 7.749 | 14.796 | 12.516 | May 28 | 3.801 | 13.209 | 12.489 | May 30 | 5.486 | 12.936 | 12.936 |
| May 21 | 5.347 | 14.704 | 12.424 | May 21 | 7.800 | 14.799 | 12.519 | May 28 | 3.905 | 13.251 | 12.531 | May 30 | 5.522 | 12.942 | 12.942 |
| May 21 | 5.383 | 14.699 | 12.419 | May 21 | 7.827 | 14.783 | 12.503 | May 28 | 3.916 | 13.247 | 12.527 | May 30 | 5.561 | 12.927 | 12.927 |
| May 21 | 5.421 | 14.694 | 12.414 | May 21 | 7.863 | 14.804 | 12.524 | May 28 | 3.927 | 13.253 | 12.533 | May 30 | 5.600 | 12.923 | 12.923 |
| May 21 | 5.446 | 14.705 | 12.425 | May 21 | 7.901 | 14.797 | 12.517 | May 28 | 3.938 | 13.255 | 12.535 | May 30 | 5.638 | 12.905 | 12.905 |
| May 21 | 5.470 | 14.704 | 12.424 | May 21 | 7.939 | 14.803 | 12.523 | May 28 | 3.948 | 13.262 | 12.542 | May 30 | 5.675 | 12.896 | 12.896 |
| May 21 | 5.511 | 14.711 | 12.431 | May 21 | 7.976 | 14.823 | 12.543 | May 28 | 4.053 | 13.297 | 12.577 | May 30 | 5.713 | 12.884 | 12.884 |
| May 21 | 5.547 | 14.710 | 12.430 | May 21 | 8.019 | 14.810 | 12.530 | May 28 | 4.063 | 13.336 | 12.616 | May 30 | 5.750 | 12.862 | 12.862 |
| May 21 | 5.585 | 14.704 | 12.424 | May 21 | 8.057 | 14.803 | 12.523 | May 28 | 4.075 | 13.340 | 12.620 | May 30 | 5.789 | 12.851 | 12.851 |
| May 21 | 5.622 | 14.726 | 12.446 | May 21 | 8.093 | 14.837 | 12.557 | May 28 | 4.178 | 13.471 | 12.751 | | | | |

[a] Magnitudes extracted from an projected radius of 312 km; we estimate that the nucleus accounted for 50-70% of the signal. The m_a values are uncorrected, while the m_c are corrected for geometric effects and adjusted for non-photometric conditions.